# Projectile's mass, reactivity and molecular dependence on ion nanostructuring

S.Bhattacharjee[1], P. Karmakar[2*], A.K. Sinha[1], A. Chakrabarti[2]

[1]UGC-DAE Consortium of Scientific Research, III/LB-8, Saltlake, Kolkata -700098, India

[2]Variable Energy Cyclotron Center, I/AF, Bidhannagar, Kolkata -700064, India

## Abstract

We have reported the dependence of projectile mass along with the chemical reactivity and nonlinear effects on ion beam induced nano structure formation when 8 keV $He^{1+}$, $N^{1+}$, $O^{1+}$, $Ar^{1+}$ atomic ions and 16 keV $N_2^{1+}$ and $O_2^{1+}$ molecular ions are bombarded on the Si (100) surface at an incidence angle of $60^0$. Ex situ atomic force microcopy (AFM) measurements reveals the ripple structure development of various forms and dimensions depending on the projectiles mass, chemical reactivity and molecular state. This experimental study explores the necessary requirements for ion induced nanopatterning and their control.

* Corresponding author, E-mail: prasantak@vecc.gov.in

# Introduction

Development of periodic ripple morphology on solid surfaces by oblique incident ion bombardment has become a subject of intense research in recent years because of the controllable wavelength and amplitude of self organized nano patterns [1] [2] . This method offers the possibility of producing large-area nano-structured surfaces and has been believed to be an economical and efficient technology for nanostructuring of surfaces.

Formation of ion induced ripple topography depends on the growth or decay of the perturbations present at the initial stage of the bombardment. The BH model [3] based on Sigmund's sputtering [4] assumes an ellipsoidal shape of the collision cascade. If the surface has a local curvature, more energy from the collision cascades reaches to the valleys than the hills and therefore the preferential sputtering of the valleys generates instability. The instability combining with thermal diffusion forms the nanostructures. The presence of local surface curvature is essential to initiate the curvature dependent instability. In case of a flat surface the ion beam itself generates the initial random rough surface due to its stochastic nature. The presence of initial roughness, impurities and preferential sputtering aids to initiate the structure formation process [5],[6,7]. In the case of semiconducting surfaces the development of the nanostructure is due to sputtering, depending on the local incident angle and surface curvature which result in surface roughness and radiation stimulated surface diffusion smoothing away topography inhomogenities [3]. The erosion rate of ion bombarded surfaces is characterized by the sputtering yield, defined as the average number of atoms leaving the surface of a solid per incident particle. In the process of sputtering the incoming ions penetrate the surface and transfer their kinetic energy to the atoms of the sample by inducing cascades of collision where most of the sputtered atoms are located on the surfaces and the scattering process which can lead to sputtering take place within a certain layer of average depth [4]. This depth depends on the energy of the incident ions. In the case of bombardment with different projectiles of same energy, it depends on the mass and molecular state of projectile ions. Lower will the mass of the ions higher will be its penetration depth. For the case of inert projectiles, it will simply sit inside the sample depending on its penetration depth without changing the chemical composition of the sample [8] where as reactive projectile changes the chemical environment during bombardment. There is the correlation between surface composition and sputtering as sputtering yield changes depending on the composition of the surfaces [9]. The development of the nanostructure might be dependent on the molecular state as the molecular ions is considered as two atomic ions sputtering the sample surface simultaneously results in increasing sputtering yield. Therefore, it is important to study the parameters other than beam energy, fluence and incident angle which change the sputtering and thereby surface topography.

In this work we have studied the effect of mass of the ions on well known and well developed ripple formation on Si (100) surfaces along with the chemical effect and the nonlinear effect of ions. We have bombarded Si (100) with the ions of same energy of 8 keV having different mass with constant fluence. We observed that the formation of ripple structures depends not only on the energy and fluence but also on the chemical reactivity. The understanding of the ripple formation nature is necessary to control this process and to use it for practical applications. Along with this the nonlinear effect of ion bombardment is also explained.

**Experimental**

Degreased and cleaned Si(100) wafers are bombarded with mass analyzed 8 keV $He^{1+}$, $N^{1+}$, $O^{1+}$, $Ar^{1+}$ atomic ions and 16 keV $N_2^{1+}$ and $O_2^{1+}$ molecular ions at a $60^0$ angle of ion incidence with the surface normal. The ion fluence was $1\times10^{18}$ ions /cm$^2$ and $2\times10^{18}$ ions/cm$^2$ for each ions, measured using a current integrator after the suppression of secondary electron emission. The effective energy for both atomic and molecular ions is same as it is assumed that the molecular beam sputters the surface as two atomic ions of half the incident energy of the molecule. The ion beam was extracted from 6.4 GHz ECR ion source of the Radioactive Ion Beam Facility at Variable Energy Cyclotron Centre Kolkata [10]. For the quantitative morphological analysis the samples were investigated in air by atomic force microscope (AFM) using Nanoscope E from Digital Instruments in contact mode.

**Results and Discussions**

AFM images of the samples bombarded at an incidence angle of $60^0$ with 8 keV $He^{1+}$, $N^{1+}$, $O^{1+}$, and $Ar^{1+}$ ions for the ion fluence of $1\times10^{18}$ ions/cm$^2$ are shown in Fig 1(a), (b), (c), (d) respectively and Fig.1(e), (f), (g), and (h) shows the AFM images of the same ions $He^{1+}$, $N^{1+}$, $O^{1+}$, and $Ar^{1+}$ respectively at same energy and incidence angle for the ion fluence of $2\times10^{18}$ ions/cm$^2$. Figures show that the nano structures developed on the Si (100) samples are more enhanced when the ion fluence is increased to $2\times10^{18}$ ions/cm$^2$ but for the same fluence the developed structures are not equivalent for all the ions of same energy because of the difference in their masses. The rms roughness is a measure of height amplitude obtained from the in built software of the AFM instrument [11] and the ripple wavelength is defined as the lateral distance between two ripples [12].

The variation of the ripple wavelength with mass and molecular state of the projectile is shown in Fig.2. The longitudinal range of different projectiles of same energy are calculated using TRIM [13]. The ripple wavelength varies linearly with the ion range and inversely with the mass of the ions. It has been shown earlier that the variation of wavelength is linear with the energy of the ions [14] and is dependent on the ion range, $a$ following an empirical relation $l=40a$ [15] where $l$ is the ripple wavelength. Our observation of inverse variation of the ripple wavelength with projectile mass (linear with ion range) is consistent with earlier data and as well as it illustrates the mass dependence.

Fig.3 illustrates the dependence of the rms roughness with mass and molecular state of the projectile. Initially with decrease in mass (increase of longitudinal range) of ions rms roughness increases linearly and gets saturated for lighter ions having mass less than nitrogen. The ion sputtering is generally determined by atomic processes taking place along a finite penetration depth inside the bombarded material. The incoming ions penetrate the surface and transfer their kinetic energy to the atoms of the substrate atoms whereas most of the sputtered atoms are located at the surface, therefore, the scattering events that might lead to sputtering takes place within a certain layer at a depth. But as the penetration depth is higher in the case of lighter ions, the collision cascades are formed deep inside the sample leading to less sputtering of surface atoms and thus roughness of the surface gets saturated.

The development of the nanostructure also depends on the reactivity of the projectile ions with the sample. When Si(100) is bombarded with the projectiles like $N^{1+}$ and $O^{1+}$, the initial bombardment develops different phases on the sample surface due to the reactive property of the projectiles ions. Further bombardment leads to the development of nanostructure due to the non uniform sputtering of the same surface because of their compositional change [8].

Once the structure are formed, local ion impact angle on the beam facing surface of the ripple is reduced resulting in the increase of implantation of the bombarding ion on the beam facing side of the ripple. This again changes the composition of the surface resulting in difference in the sputtering yield between the two phases of the same surface. In case of bombardment of oxygen and nitrogen ions on Si(100) the ripple formation is well-defined compared to argon and helium ions. Due to higher reactivity of the $N^{1+}$ and $O^{1+}$ ions there forms $Si_xN_y$ and $SiO_z$ respectively with the initial ion bombardment and thus due to change in the chemical composition of the surface there develops a difference in the sputtering yield on the sample surface resulting in quick development of ripple structure due to preferential sputtering [8,9]. The altered layer of silicon irradiated with 10keV $O_2^+$ ions for $\theta > 20^0$, $SiO_x$ are formed [16]. Similarly FTIRS study of superficial Si layers formed by 9 keV $N_2^+$ ion bombardment at impact angles up to $70^0$ demonstrated the existence of $Si_3N_4$ absorption centers [17]. $Ar^+$ forms a damaged layer with the original composition of Si (100).Argon being an inert gas the retention of Ar on Si is much lower as compared to oxygen [8] and nitrogen [9] hence the ripple structure formation is not possible with this combination of fluence and energy. With further increase of either energy or fluence it is possible to developed ripple structure with the argon ion. But simultaneously helium being an inert gas the nano structure formation is possible due to its lower mass. With an advantage of blister formation $He^{1+}$ ion could develop nanostructure on Si(100) [18]. Earlier Y. Yamauchi et al showed the formation of bubble structure on silicon by helium ion bombardment. At room temperature bubbles of 200 nm radius were formed and also the retention of He with silicon increases with the ion fluences at room temperature [19].

Fig. 4(c) and (d) shows the AFM images of the samples bombarded with 16kV $N_2^{1+}$ and $O_2^{1+}$ molecular ions respectively at an incidence angle of $60^0$. Fig. 4(a) and (b) shows the same for 8 kV $N^{1+}$ and $O^{1+}$ ions at same fluence of $1\times10^{18}$ ions/cm$^2$ respectively. In the case of molecular ions as two atomic ions sputters the surface at same instant the nano structure development is faster in this case. Comparing the figures for the case of nitrogen and oxygen molecular and atomic ions it is clear that the well defined ripple structures are developed at lower fluence for the molecular ions. With further increase of ion fluence the rms roughness of the surface has increased.

Fig.5 shows the variation of ripple wavelength and surface roughness with the ions mass. The ripple wavelength and roughness both shows a decrease in their values with increasing mass of the atomic ions but in the case of molecular ions the value of the ripple wavelength and roughness are high with respect to their respective values in the case of atomic ions. It is due to the overlapping of the collision cascade for cluster ion bombardment, resulting in higher sputtering yield as it was assumed that the molecular beam sputters the surface as two atomic ions of half the incident energy of the molecule at the same instant. Experimentally the nonlinear effect has been reported in the case of $Ar^+$ bombarded Si (100) [20] and graphite (HOPG) surfaces [21] . Also among $N_2$ and $O_2$ the effect of $O_2$ is higher as that of nitrogen which forms a stable structure whereas oxygen under goes increasing disturbance during growth until the final destruction of the pattern and a possible origin of this physical effect is the high diffusivity of excess oxygen in $SiO_2$ as compared to the low diffusivity of excess nitrogen in $Si_3N_4$. Thus, nitrogen ion beams could form more promising ripple pattern [22].

In conclusions, the formation of ripple structure is dependent on incident ion energy, fluence, chemical reactivity and nonlinear effect.

## Acknowledgement

The authors would like to thank Dr. V. Ganesan for accessing the SPM and Mr. M. Gangrade for doing SPM measurements.

**Figure Caption**

Fig.1 AFM images of the samples bombarded with (a) $He^{1+}$, (b) $N^{1+}$, (c) $O^{1+}$, and (d) $Ar^{1+}$ ions at an angle of $60^0$ for the ion fluence of $1\times10^{18}$ ions/cm$^2$ and Fig.1 (e), (f), (g), and (h) shows the AFM images for the same ions $He^{1+}$, $N^{1+}$, $O^{1+}$, and $Ar^{1+}$ respectively at an ion fluence of $2\times10^{18}$ ions/cm$^2$.

Fig.2 The variation of the ripple wavelength with the longitudinal range.

Fig.3 The variation of the rms roughness with the longitudinal range.

Fig.4 AFM images of the samples bombarded with (a) $N^{1+}$ and (b) $O^{1+}$ atomic ions at an angle of $60^0$ with the ion fluence of $2\times10^{18}$ ions/cm$^2$. Fig.4 (c) and (d) shows the same for the $N_2^{1+}$ and $O_2^{1+}$ molecular ions at the fluence of $2\times10^{18}$ ions/cm$^2$ respectively.

Fig.5 The variation of ripple wavelength and surface roughness with the ions.

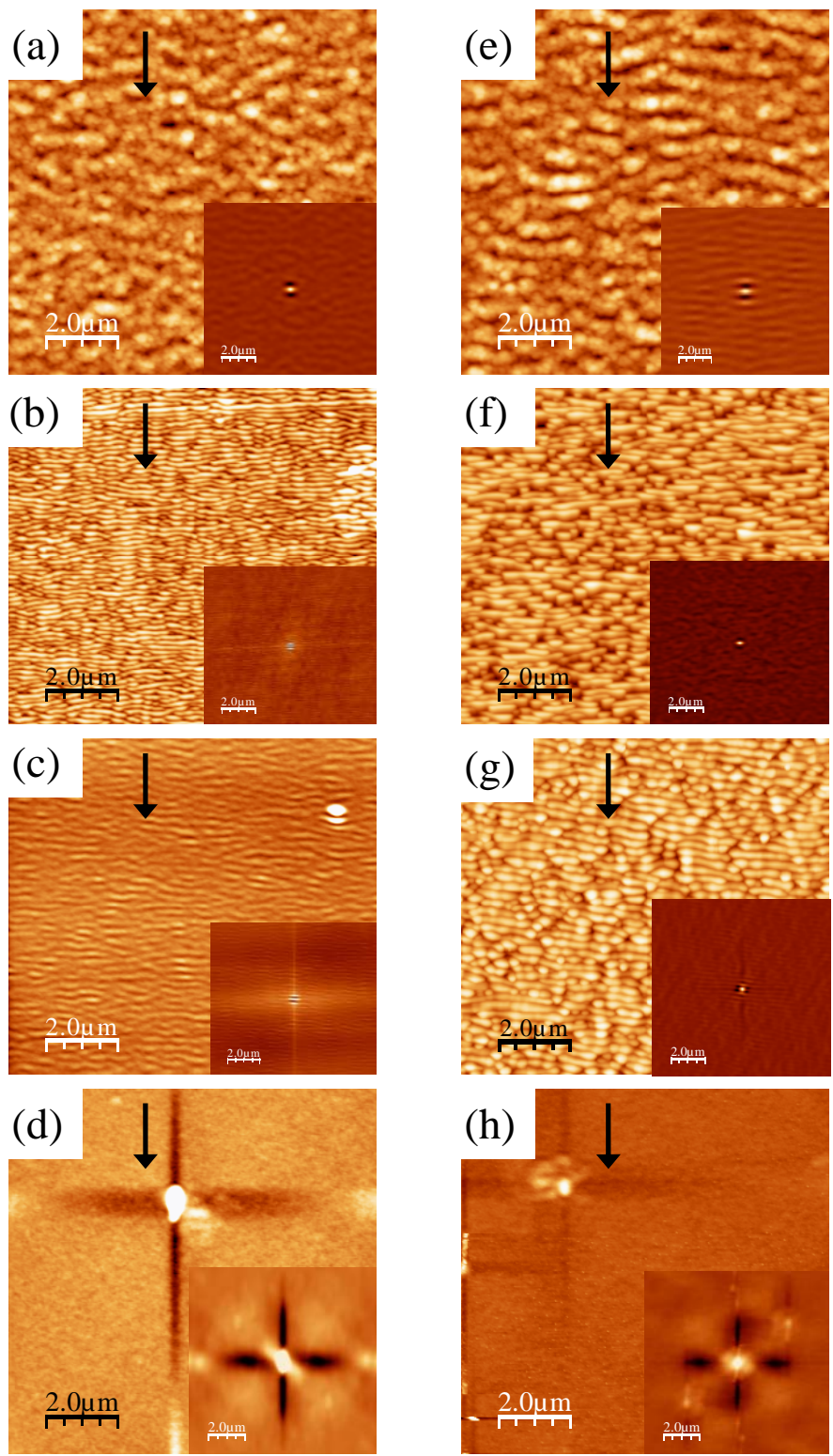

**Figure 1**

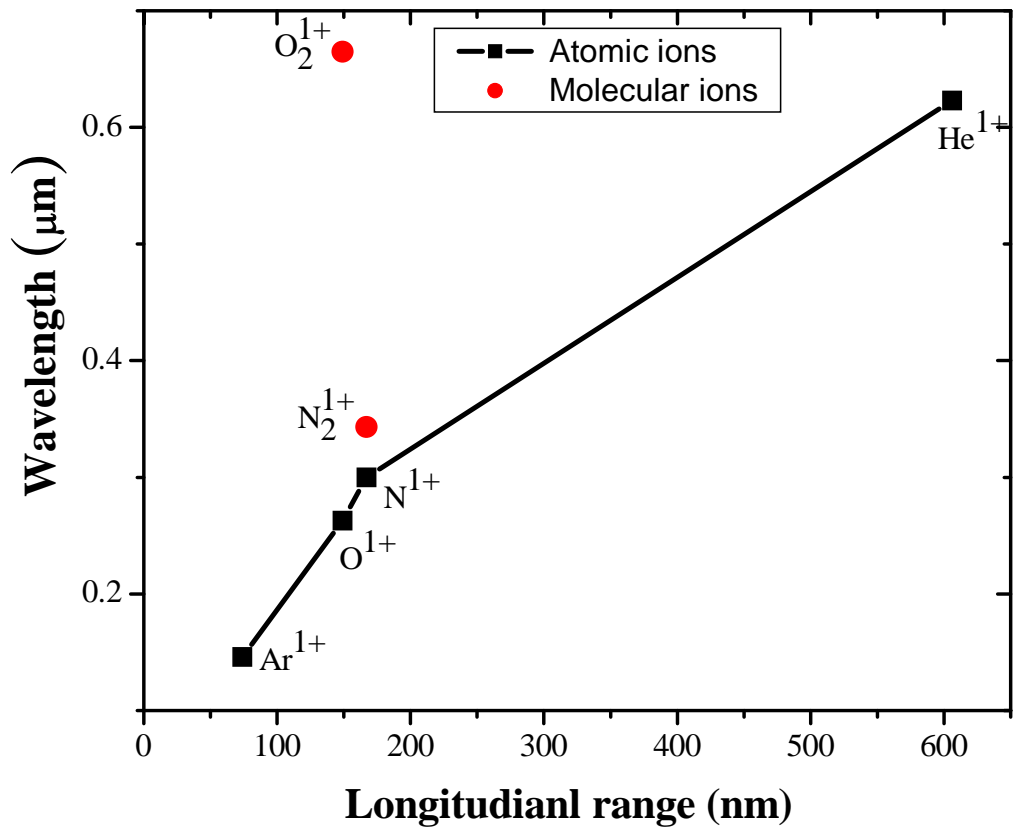

**Figure 2**

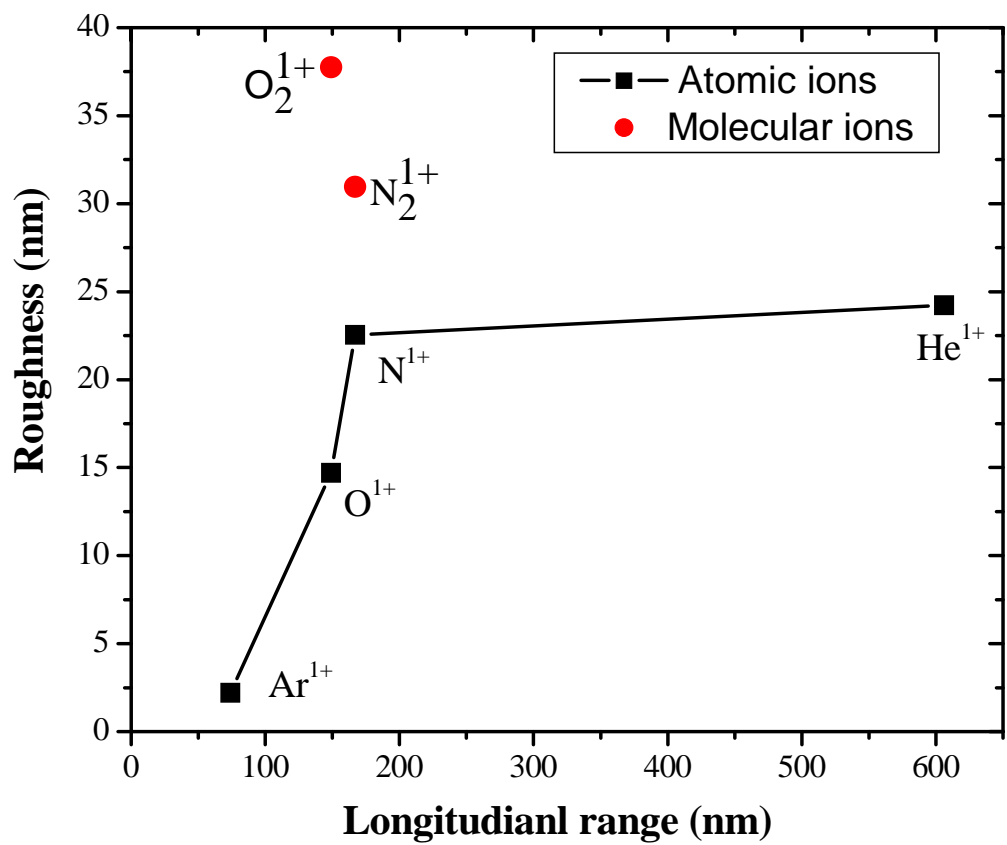

**Figure 3**

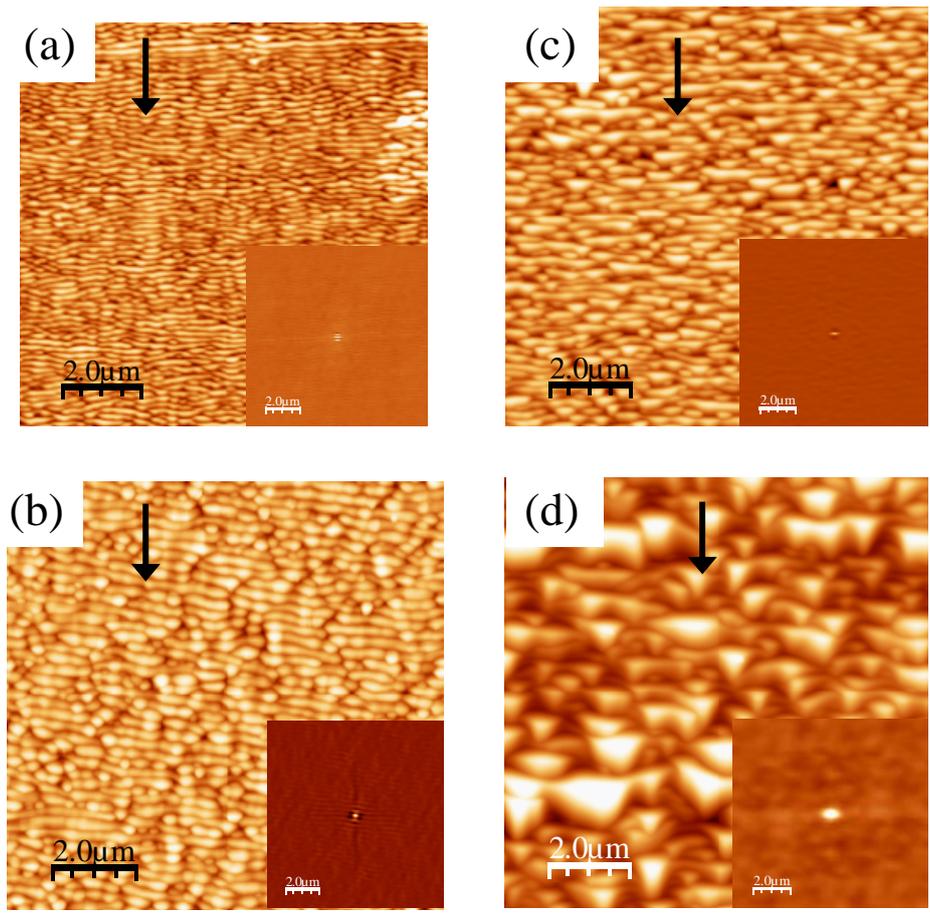

**Figure 4**

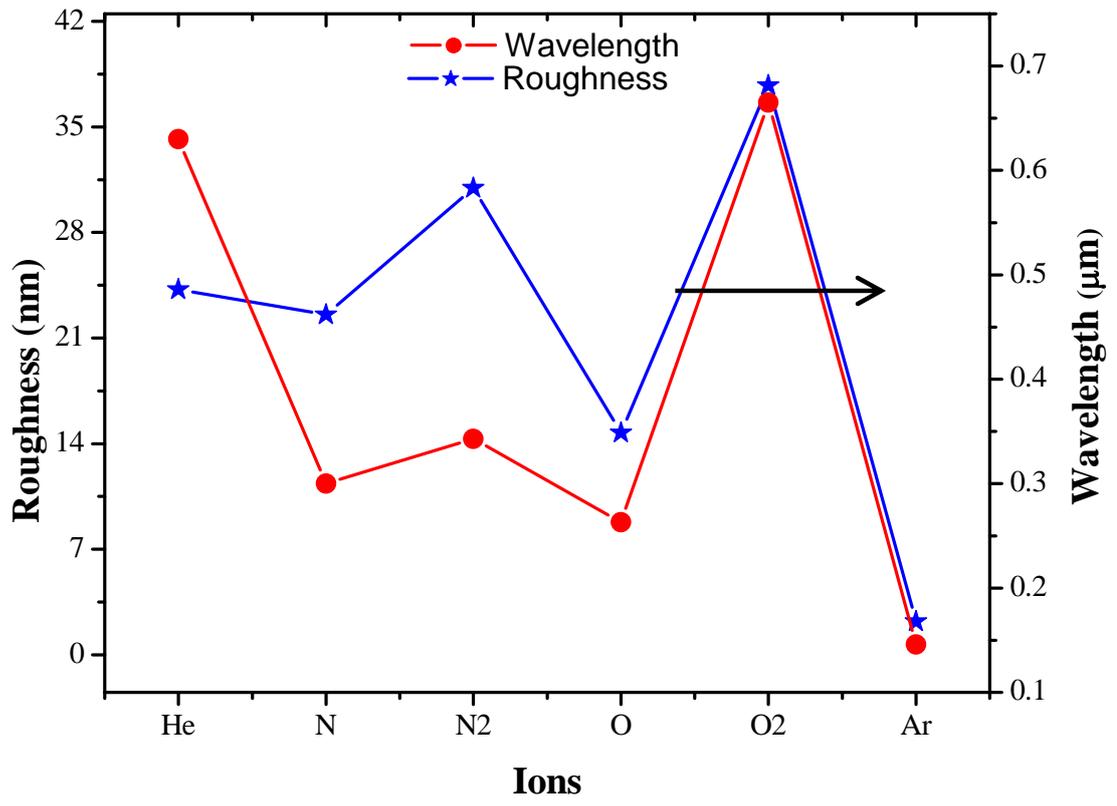

**Figure 5**